%% file: main.tex
\documentclass{vgtc}                          

\graphicspath{{figures/}{pictures/}{images/}{./}} 

\usepackage{times}                     

\usepackage{xcolor}

\usepackage{float}

\usepackage{graphicx}
\usepackage{tabularx}
\usepackage{array}
\usepackage{longtable}
\usepackage{wrapfig}
\usepackage{caption}
\newcolumntype{C}{>{\centering\arraybackslash}X}
\usepackage{booktabs, multirow} 
\usepackage{xcolor,colortbl} 
\usepackage{changepage,threeparttable} 
\usepackage{hyperref}
\usepackage{mathptmx}                  

\onlineid{0}

\vgtccategory{Research}

\title{Quick-View Takeaways: \\ How Does Title Framing Influences Pattern Identification in Line Charts?}

\author{Jasmine Lim\thanks{e-mail: \{jasmine.t.lim, pandey, quadri\}@ou.edu} \\ %
        \scriptsize University of Oklahoma %
\and Tapendra Pandey\footnotemark[1] \\ %
     \scriptsize University of Oklahoma %
\and Arran Zeyu Wang\thanks{e-mail: zeyuwang@cs.unc.edu} \\ %
     \scriptsize University of North Carolina at Chapel Hill %
\and Ghulam Jilani Quadri\footnotemark[1] \\ %
     \scriptsize University of Oklahoma}

\abstract{
    Visual data communication in digital media is increasingly characterized by short attention spans and snapshot-based viewing, often employing line charts to convey trends and patterns.
   Among all visual elements, titles are 
   crucial ones that can shape how viewers interpret visual information and form chart takeaways.
   In this study, we examine how title characteristics, particularly title word count and intended message, influence people's pattern identification in single-class line charts.
   Participants viewed 50 line charts collected from online news media and identified the pattern they perceived.
   Our results demonstrate that both title word count and intended message significantly influence viewers' pattern identification.
   Our findings highlight the importance of title design in shaping chart takeaways and 
   effective visualization communication.

} 

\begin{document}


\firstsection{Introduction}
\maketitle
\enlargethispage{0.5in}

\input{sections/1-intro}
\label{sec-intro}
\section{Study Design}
\input{sections/3-experiment}
\label{sec-experiment}
\vspace{-5pt}
\section{Analysis and Results}
\input{sections/4-results}
\label{sec-results}
\vspace{-5pt}

\section{Discussion}
\input{sections/5-discussion}
\label{sec-discussion}

\bibliographystyle{abbrv-doi}

\bibliography{bibliography}
\end{document}

%% file: sections/1-intro.tex
Line charts are commonly used in online news media to communicate trends, comparisons, and changes over time \cite{quadri2021survey,franconeri21}. Alongside visual structure, titles provide readers with contextual information that can frame how a chart is understood \cite{kim2021understandingReadersIntegrateChartsCaptions}
. 
Readers often form quick judgments on line charts by identifying high-level patterns such as trends, peaks, and declines
guiding viewers to interpret the visualized data. As the primary textual feature accompanying many news charts, titles can influence how viewers understand the visualized information~\cite{stokes2022strikingBalanceIntegratingTextChart}. Therefore, understanding the role of title framing in pattern identification is significant for effective visual communication.


%
Titles vary in the messages they communicate and amount of text they contain, representing two key dimensions of variation commonly observed in real-world news visualizations: title word count and intended message.
Prior work examined individual aspects of textual framing using controlled chart stimuli and found that textual cues can influence comprehension, recall, judgment, and chart takeaways \cite{kong2018slantsTitles,stokes2022strikingBalanceIntegratingTextChart, kim2021understandingReadersIntegrateChartsCaptions}. In addition, textual framing can communicate different messages, ranging from descriptions of visible patterns to broader contextual explanations and insights \cite{stokes2022strikingBalanceIntegratingTextChart}. However, it remains unclear how title characteristics influence viewers' pattern identification during quick viewing of real-world news charts.

In this work, we investigate how two characteristics of title framing: intended message and title word count, influence pattern identification in single-class line charts collected from online news media. Our stimuli contain titles that communicate neutral, statistical, and affective messages, identified from the content of articles surrounding the chart~\cite{kong2018slantsTitles}.
We conducted a user study with 47 participants, who viewed 50 charts and identified the pattern from seven predefined categories (see \autoref{fig:stimuli}b) 
to investigate how title framing in single-class line charts affects readers' pattern takeaways.
We found that pattern identification varied significantly across both dimensions of title framing: word count and intended message.
However, a visually salient structure helps the viewer identify the pattern, even when the title framing remains neutral and short. We further observe 
variability across charts, suggesting that textual framing and salient visual structure jointly contribute to the pattern 
identification.
Our findings provide empirical evidence on the role of title framing in shaping chart takeaways and offer guidance for designing public-facing visualizations that align textual cues with designers' communicative objectives.

%% file: sections/3-experiment.tex
\begin{figure*}[t]
  \centering
  \includegraphics[width=\textwidth]{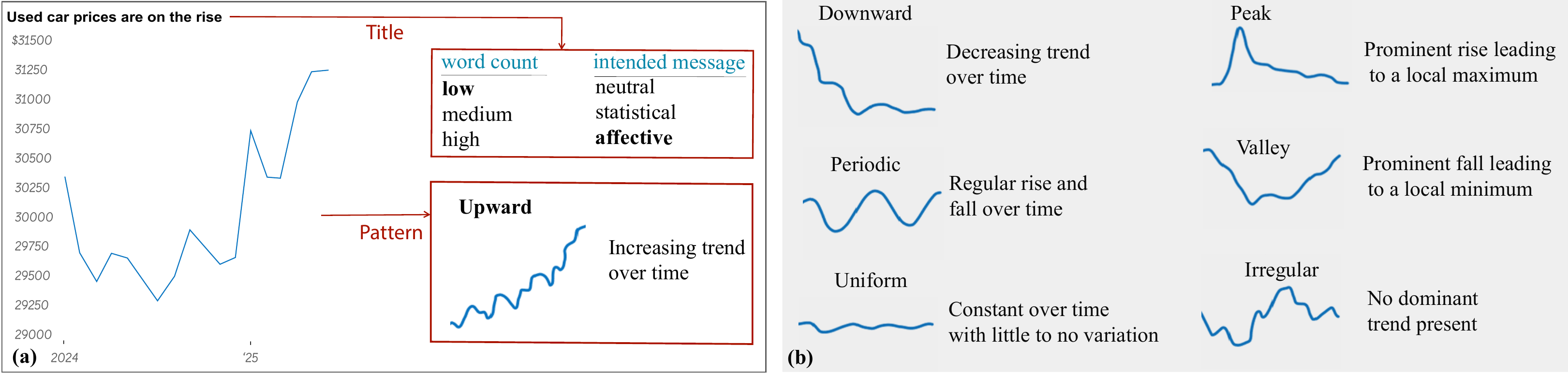}
  \vspace{-20pt}
  \caption{Left (a) shows the example stimuli used in our study, with the title word count -- \textbf{low}, intended message -- \textbf{affective}, and intended pattern -- \textbf{upward}. Right (b) shows the remaining intended patterns along with their definitions. }
  \vspace{-10pt}
  \label{fig:trend}
\end{figure*}

\noindent\textbf{Stimuli:} We collected 50 single-class static line charts 
from \href{nytimes.com}{The New York Times} and \href{wsj.com}{Wall Street Journal}
that have a visible title and a clearly identifiable statistical pattern.
To capture common patterns observed in real-world news visualizations and provide a consistent 
vocabulary for pattern identification, we defined seven possible intended patterns (\textit{upward, downward, peak, valley, periodic, uniform}, and \textit{irregular}), as shown in \autoref{fig:trend}b.

\noindent\textbf{Title Framing:} 
%
We characterize each chart title along two textual framing dimensions: title word count and intended message, as illustrated in \autoref{fig:trend}a. Title word count refers to the number of words in the title of a given chart, including prepositions, but excluding symbols. We categorized the title word count as low (1-5 words), medium (6-10 words), and high (11+ words). The intended message represents the designer’s intended framing of how the data and pattern should be interpreted. We categorized the intended message as \textit{neutral}, \textit{statistical}, or \textit{affective} \cite{kong2018slantsTitles}. \textit{Neutral} titles describe the topic with minimal interpretation. \textit{Statistical} titles emphasize numeric values, magnitudes, or quantitative comparisons. \textit{Affective} titles contain evaluative or emotionally suggestive language without any statistics.
%
Next, we identified the intended pattern for each line graph by analyzing its visual structure alongside the corresponding article's title, caption, and content, treating the pattern most strongly conveyed by both the visualization and the narrative as the ground truth for evaluating participants' identified patterns.
The intended pattern (as shown in~\autoref{fig:trend}) served as the ground truth for evaluating the correctness of participants' identified patterns.

\noindent\textbf{Procedure:} 
Following a power analysis (N=28), 
we recruited 47 participants, including 22 via university mailing lists and online announcements and 25 via Prolific.
Participants were compensated at a rate of \$12 per hour.
We conducted a web-based study where participants viewed one chart at a time and identified the pattern they perceived. Participants reported their age and educational background.
Before starting the study, participants reviewed the definitions of all seven predefined patterns. Each participant completed 50 trials. In each trial, a chart was displayed for up to 15 seconds while participants answered the question: “What pattern can you see in the line chart?” Participants selected one pattern label from the multiple-choice options using radio buttons.
We included two attention check trials across the study to eliminate random clicks.

%% file: sections/4-results.tex
To examine whether title framing influences pattern identification, we analyzed the title word count and the intended message using both quantitative and qualitative analyses. We excluded
5 participants who failed at least one attention check ($n=2$) or reported color blindness ($n=3$), leaving a final sample of 42 participants.
Across 42 participants and 2100 responses, we excluded an additional 13 responses due to timeouts, resulting in 2087 valid responses.
%
Two independent coders assigned an intended pattern label to each chart using the predefined pattern categories; inter-rater reliability was 
substantial (Cohen's $\kappa$ = 0.85); we resolved disagreements through discussion to reach consensus. \autoref{tab:50} shows the distribution of the 50 single-class line charts coded into framing factors and intended pattern.

\renewcommand{\arraystretch}{0.9}
\begin{table}[!h]\centering
\vspace{-5pt}

\caption{Distribution of 50 single-class line charts from our study.
}
{\footnotesize
\setlength{\tabcolsep}{3.5pt}
\renewcommand{\arraystretch}{1.0}
\vspace{-5pt}
\begin{tabular}{
p{1.0cm}
p{0.4cm} |
p{1.2cm}
p{0.4cm} |
p{3.8cm}
}
\hline

\multicolumn{1}{>{\raggedright\arraybackslash}m{1.0cm}}{\textbf{Word Count}} &
\multicolumn{1}{>{\raggedright\arraybackslash}m{0.4cm}|}{\textbf{N}} &
\multicolumn{1}{>{\raggedright\arraybackslash}m{1.2cm}}{\textbf{Intended Message}} &
\multicolumn{1}{>{\raggedright\arraybackslash}m{0.4cm}|}{\textbf{N}} &
\multicolumn{1}{>{\raggedright\arraybackslash}m{3.8cm}}{\textbf{Intended Pattern (N)}} \\
\hline

Low    & 18 & Neutral     & 19 & Upward (12), Downward (3) \\
Medium & 22 & Statistical & 23 & Peak (12), Valley (9), Periodic (2) \\
High   & 10  & Affective   & 8  & Uniform (0), Irregular (12) \\

\hline
\end{tabular}}

\label{tab:50}
\vspace{-5pt}
\end{table}

\textbf{Log-linear Analysis:} To examine whether textual framing factors
are associated with pattern identification, we fitted a hierarchical log-linear model, which
examines relationships in multi-way contingency tables and tests whether the distribution of identified pattern labels (dependent variable) varies across levels of textual framing factors.
We compared a baseline model independent of textual framing ($\chi^2 = 11626.45$) with a model that allowed responses to vary by title word count and intended message.~The interaction model provided a significantly better fit ($\Delta \chi^2 = 268.48$, $df=28$, $p < .001$), indicating that the distribution of pattern labels differs across framing factors. This model treats each response as independent, although responses are nested within 42 participants and 50 charts; we therefore report effect sizes.

Both title word count ($\Delta\chi^2 = 155.06$, $df = 14$, $p < .001$, Cramér's $V = .19$) and intended message ($\Delta\chi^2 = 113.42$, $df = 14$, $p < .001$, {Cramér's $V = .16$}) independently improved the model fit. 
%
{We also examined pattern agreement across levels of the textual framing factors by comparing }
the designers' intended pattern and the participants' identified pattern. 
Our qualitative analysis showed higher agreement (i.e., more than 50\% of participants' identified patterns aligned with the designers' intended pattern) for titles
with an affective intended message (50.0\%, 6/12 charts) and high title word counts (88.9\%, 8/9 charts).

%% file: sections/5-discussion.tex
\noindent \textbf{Title framing may guide pattern takeaways:} 
%
%
%
%
Our findings show that pattern identification varied across title framing factors in single-class line charts. Charts with affective titles and higher title word counts exhibited greater agreement with the intended pattern. These findings suggest that chart titles may provide interpretive cues that help viewers in identifying visual patterns. Consequently, the design and phrasing of chart titles play a role in visualization communication by shaping viewers' interpretations and influencing the takeaways they derive from a chart.

\begin{figure}
     \centering
     \includegraphics[width=\columnwidth,height=0.25\textheight,keepaspectratio]{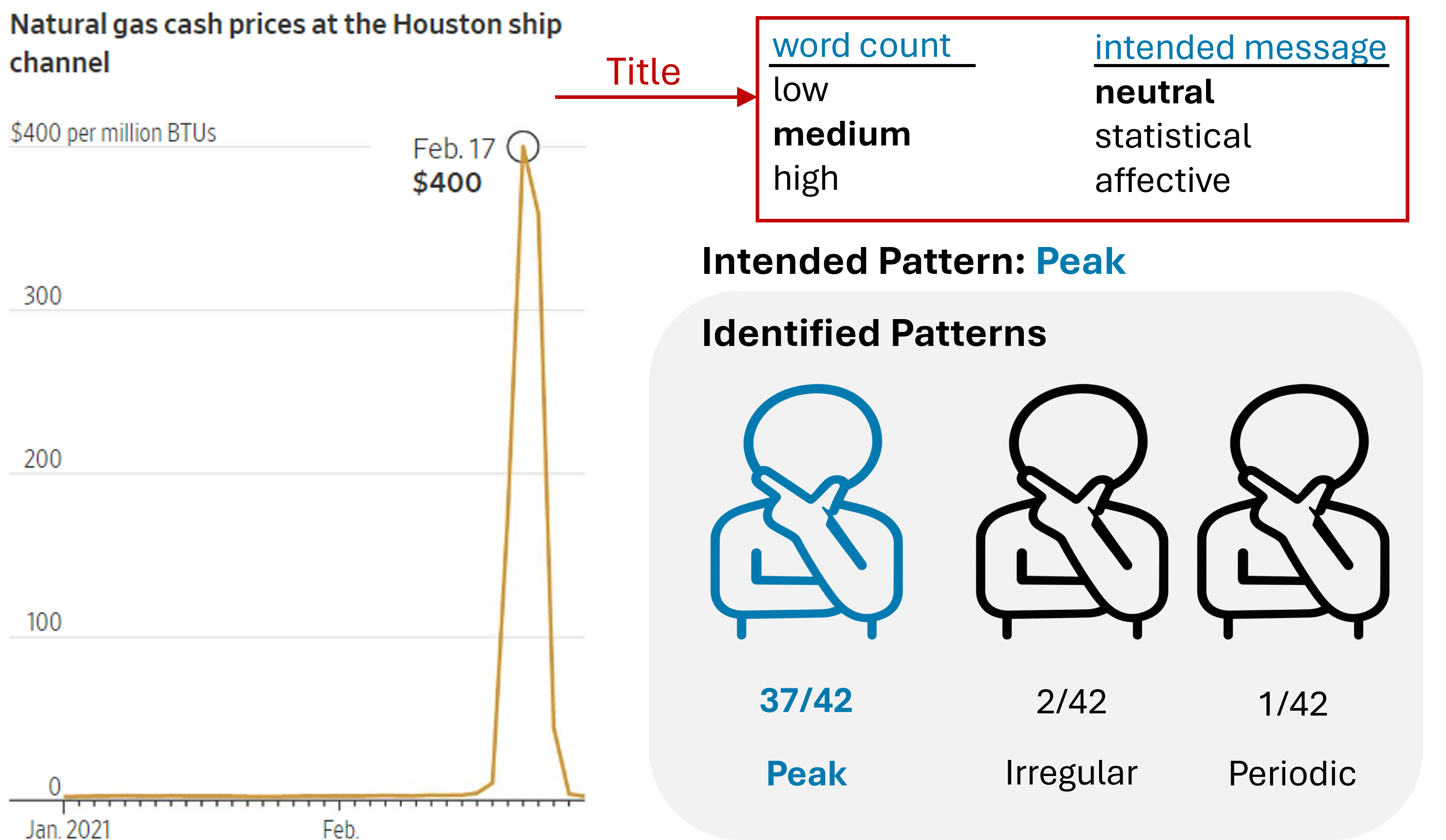}
      \caption{Example chart showing a visually salient peak pattern.}
      \label{fig:stimuli}
      \vspace{-20pt}
\end{figure}

\noindent \textbf{Salient Visual Structure:}
Our findings suggest that salient visual structures may support pattern identification. Many line charts in our stimulus pool contained visually prominent patterns, and participants frequently identified patterns that aligned with these structural characteristics.
%
For example, \autoref{fig:stimuli} features a prominent peak pattern and a neutral title that offers little interpretive guidance, suggesting that sufficiently salient visual patterns can support pattern identification even in the absence of strong textual cues.
However, not all charts emphasize a single dominant pattern. Some support multiple plausible interpretations, making the intended pattern less apparent from visual structure alone. In these cases, chart titles may provide additional interpretive cues. 

Considered together, our observations suggest that pattern identification may be shaped by both salient visual structure and title framing, not by either factor in isolation.

\noindent \textbf{Future Work:} 
%
Future work should investigate a broader range of textual characteristics, including text type, quantity, and context, through systematic empirical studies with balanced stimulus sets. This work could establish a framework for understanding how viewers integrate textual information when interpreting charts and inform design guidelines for textual elements that better support the intended chart takeaways.